\documentclass[12pt, fleqn]{article}
\usepackage[cp1251]{inputenc}
\usepackage{latexsym,amsfonts,amssymb}
\usepackage{graphicx}

\usepackage{amsbsy}
\usepackage{amsmath}
\usepackage{epsf}

\newtheorem{theo}{Theorem}
\newcommand{\bt}{\begin{theo}}
\newcommand{\et}{\end{theo}}
\newcommand{\bd}{\begin{displaymath}}
\newcommand{\ed}{\end{displaymath}}

\newcommand{\be} {\begin{equation}}
\newcommand{\ee} {\end{equation}}
\newcommand{\ba} {\begin{array}}
\newcommand{\ea} {\end{array}}

\newcommand{\p} {\partial}

\newcommand{\lbd} {\lambda}

\begin{document}
%%\normalsize
 \begin{center}
 {\Large \bf Conditional symmetries
 for systems of PDEs: \\ new definitions and their application \\ for
 reaction-diffusion systems }\\
 \medskip

{\bf Roman Cherniha$^{\dag,}{^\ddag}$}
%%\footnote{\small e-mail: cherniha@imath.kiev.ua}
 \\
{\it  $^\dag$~Institute of Mathematics, Ukrainian National Academy
of Sciences,\\
 3 Tereshchenkivs'ka Street, Kyiv 01601, Ukraine\\
  $^\ddag$~Department  of  Mathematics,
     National University
     `Kyiv-Mohyla Academy', \\ 2 Skovoroda Street,
     Kyiv  04070 ,  Ukraine
 }\\
 \medskip
 E-mail: cherniha@imath.kiev.ua
\end{center}

\begin{abstract}

New  definitions of $Q$-conditional
  symmetry for systems of PDEs  are presented,
  which generalize the standard notation of
  non-classical (conditional) symmetry.
  It is shown that different types of $Q$-conditional
  symmetry of a system  generate  a  hierarchy  of conditional symmetry
  operators. A class of  two-component  nonlinear reaction-diffusion systems
   is examined  to demonstrate applicability of the definitions
   proposed and it is shown when  different definitions
   of $Q$-conditional symmetry lead to the same operators.

\end{abstract}

\begin{center}
{\bf 1.Introduction}
\end{center}

Since 1952, when A.C. Turing published the remarkable paper
\cite{turing}, in which  a revolutionary idea about mechanism of
morphogenesis (the development of structures in an organism during
its life) has been proposed,
 nonlinear reaction-diffusion  %@@
systems have been extensively studied by means of different
mathematical methods, including group-theoretical methods (see
\cite{ch-king,ch-king2,niki-05,ch-king4} and the papers cited
therein). The main attention was paid to investigation of
 the two-component RD systems of the form
 \be\label{1*}\ba{l}
 U_t=[D^1(U)U_x]_x+F(U,V),\\
V_t=[D^2(V)V_x]_x+ G(U,V)
 \ea\ee
where
  $U= U(t,x)$ and $V= V(t,x)$ are two  unknown functions representing the  densities
  of populations (cells, chemicals etc.),
$F(U,V)$ and $G(U,V)$ are two given functions describing interaction
between them and environment,
 the functions $D^1(U)$ and $D^2(V)$ are the relevant diffusivities
 %%(usually they are to be constants or power functions),
 (hereafter they are assumed positive functions)  and the subscripts $t$ and $x$ denote
differentiation with respect to these variables.
Note that   nonlinear system (\ref{1*}) generalizes many well-known nonlinear  %@@
second-order models used to describe  %@@
various processes in physics \cite{ames},   %@@
biology \cite{mur2, britton} and ecology \cite{okubo}.

At  present, one can claim that all possible  Lie symmetries of
(\ref{1*}) with the constant diffusivities are completely described
in \cite{ch-king,ch-king2}, while  it has been done for  the
non-constant diffusivities in \cite{ch-king4}. However, the problem
 of construction of $Q$-conditional symmetries (non-classical symmetries)
  for (\ref{1*})  is  still not
  solved.
  Moreover, to  our best knowledge, there are only a few papers
  devoted to the search of conditional symmetries
   of  evolution  systems,
  which contain RD systems as  a subclass  \cite{lou-92} --\cite{ch-pli-08}.
  Notably, some general results about
  $Q$-conditional symmetries of  systems with power diffusivities
  of the form
 \be\label{3}\ba{l}
 U_t=(U^kU_x)_x+F(U,V),\\
V_t=(V^lV_x)_x+ G(U,V)
 \ea\ee
 have been obtained in the recent paper \cite{ch-pli-08}.
 However, the results obtained in \cite{ch-pli-08}
 cannot be adopted
 for any system of the form (\ref{1*}).
Moreover, the authors clearly indicated that two interesting cases,
 $l=k=0$  and $l=k=-1$,  wasn't studied therein.

 It should be noted that,
 starting from the  pioneering work \cite{bl-c},  many papers were devoted to the construction of  such symmetries
 for the  {\it scalar} nonlinear RD equations of the form
\cite{fss}--\cite{ch-pliu-2006} (the reader may find more references
in the  recent book \cite{bl-anco-10})

 \be\label{1a}
 U_t=\left[D(U)U_x\right]_x + F(U).
\ee and (\ref{1a}) with the  convective term   $B(U)U_x$ (here
$B(U), D(U)$ and $F(U)$ are  arbitrary smooth functions). It is
well-known that conditional symmetries can be applied for finding
exact solutions of the relevant equations, which are not obtainable
by the classical Lie method. Moreover the solutions obtained in such
a way may have   a  physical or biological interpretation (see,
e.g.,  examples in  \cite{dix-cla, broad-2004, ch-pliu-2006}). Thus,
the same should be expected in the case of RD systems.
% In this paper we will demonstrate this using the  $Q$-conditional symmetries  of non-linear RD
%systems of the form (\ref{1}).

In the present paper, we shall consider a general multi-component
system of evolution PDEs and its particular case, system (\ref{1*}).
  The paper is organized as follows.
  In the second section, we present  new  definitions of $Q$-conditional
  symmetry, which create a  hierarchy  of conditional symmetry
  operators. In the third section, the definitions are used
  to construct the systems of determining equations(DEs) for
  the RD system (\ref{1*}).  An analysis of the derived systems
  is carried out, particularly, an example is presented, which
  illustrates different types of   $Q$-conditional
  symmetries. In the fourth section, the the  definition of $Q$-conditional
  symmetry is applied to find new exact solutions of the classical Lotka-Volterra  system
  with diffusive terms.
 Finally, we present some conclusions in the last section.

\begin{center}
{\textbf{2. Definitions of conditional symmetry for systems of
PDEs}}
\end{center}

Here we  present  new  definitions of $Q$-conditional symmetry,
which naturally arise for {\it systems} of PDEs. To avoid possible
difficulties that can occur in the case of arbitrary system of PDEs,
we restrict ourself on systems of evolution PDEs (the RD system
(\ref{1*}) is a typical example).

Consider  a  system of $m$ evolution equations ($m \geq 2$) with $2$
independent $(t, x)$ and $m$ dependent $u = (u_1, u_2, \ldots, u_m)$
variables. Let us assume that the $k$-th order ($k\geq 2$) equations
of evolution type
\begin{equation}
u_t^i=F^i \left(t,x, u, u_x, \ldots , u_{x}^{(k_i)}\right), \ i =
1,2, \ldots, m, \label{4}
\end{equation}
are  defined on a domain $\Omega \subset
%\mathbb{R}
{\bf R}^2 $ of independent variables $t$ and $x$. Hereafter $F^i$
are smooth functions of the corresponding variables, the subscripts
$t$ and $x$ denote differentiation with respect to these variables,
$u_t^i= \frac{\partial u_i}{\partial t},$ and
$u_{x}^{(j)} \equiv \frac{\partial^j u}{\partial x^j}=
\left(\frac{\partial^j u_1}{\partial x^j},\ldots ,\frac{\partial^j
u_m}{\partial x^j}\right),\, j = 1,2, \ldots,k_i$.

It is well-known that to find Lie invariance  operators, one needs
to consider   system  %@@
(\ref{4}) as the manifold ${\cal{M}}=\{S_1=0,S_2=0, \dots, S_m=0 \}$,  where%@@
\begin{equation}
 S_i \equiv u_t^i-F^i \left(t,x, u, u_x, \ldots , u_{x}^{(k_i)}\right)=0, \ i =
1,2, \ldots, m, \label{5}
\end{equation}
in the prolonged space of the  variables:  %@@
\[t, x, u, \mbox{\raisebox{-1.1ex}{$\stackrel{\displaystyle  %@@
u}{\scriptstyle 1}$}},\dots, \mbox{\raisebox{-1.1ex}{$\stackrel{\displaystyle  %@@
u}{\scriptstyle k}$}}. \] Here $k =
\max \{k_i, \ i = 1, \ldots, m \}$ and the symbol $ \mbox{\raisebox{-1.1ex}{$\stackrel{\displaystyle  %@@
u}{\scriptstyle j}$}}$ ($j = 1,2, \ldots ,k$) denotes totalities of
the $j$-th order derivatives w.r.t. the variables $t$ and $x$.

According to the definition, system  (\ref{4}) is invariant (in Lie
sense !) under
the transformations generated by the  %@@
infinitesimal operator  %@@
\be\label{6}  %@@
Q = \xi^0 (t, x, u)\p_{t} + \xi^1 (t, x, u)\p_{x} +  %@@
 \eta^1(t, x, u)\p_{u_1}+\ldots+\eta^m(t, x, u)\p_{u_m},  \ee  %@@
if the following invariance conditions are satisfied:  %@@
\be\label{7}  %@@
\mbox{\raisebox{-1.6ex}{$\stackrel{\displaystyle  %@@
Q}{\scriptstyle k}$}} S_i  %@@
 \equiv  \mbox{\raisebox{-1.6ex}{$\stackrel{\displaystyle  %@@
Q}{\scriptstyle k}$}}  %@@
\left(u_t^i-F^i \left(t,x, u, u_x, \ldots ,
u_{x}^{(k_i)}\right)\right) \Big\vert_{\cal{M}}=0,  \ i = 1,2,
\ldots, m.
\ee  %@@
The operator $ \mbox{\raisebox{-1.1ex}{$\stackrel{\displaystyle  %@@
Q}{\scriptstyle k}$}} $  %@@
is the k-th order  %@@
 prolongation of the operator $Q$ and its coefficients  %@@
are expressed  via the functions $\xi^0, \xi^1, \eta^1 , \ldots ,
\eta^m$ by well-known formulae (see, e.g., \cite{ovs}-\cite{olv}).

The crucial idea used for introducing the notion of $Q$-conditional
symmetry (non-classical symmetry) is to change the manifold
${\cal{M}}$, namely: the operator $Q$  is used to reduce
${\cal{M}}$. It is important to  note that there are several
different possibilities to realize this idea in the case of systems
containing $m$ PDEs. In fact, different definitions can be
formulated for such systems depending on the number of complementary
conditions generated by the operator $Q$.
% Now we present the corresponding definitions in the case of evolution systems of the form  (\ref{4}).
%%To avoid a discussion about the proper structure

\noindent {\textbf{ Definition 1.}} Operator (\ref{6}) is called the
$Q$-conditional symmetry of the first type  for  an evolution system
of the form  (\ref{4}) if the following invariance conditions are
satisfied:
\be\label{8}  %@@
\mbox{\raisebox{-1.1ex}{$\stackrel{\displaystyle  %@@
Q}{\scriptstyle k}$}} S_i  %@@
 \equiv  \mbox{\raisebox{-1.1ex}{$\stackrel{\displaystyle  %@@
Q}{\scriptstyle k}$}}  %@@
\left(u_t^i-F^i \left(t,x, u, u_x, \ldots ,
u_{x}^{(k_i)}\right)\right)\Big\vert_{{\cal{M}}_1} =0,  \ i = 1,2,
\ldots, m, \ee
 where the manifold
${\cal{M}}_1$ is  one  $\{S_1=0,S_2=0, \dots, S_m=0, Q(u_{i_1})=0
\}$ with a fixed number $\ i_1 \, ( 1\leq i_1 \leq m)$.

\noindent {\textbf{ Definition 2.}} Operator (\ref{6}) is called the
$Q$-conditional symmetry of the $p$-th type  for  an evolution
system of the form  (\ref{4}) if the following invariance conditions
are satisfied:
\be\label{9}  %@@
\mbox{\raisebox{-1.1ex}{$\stackrel{\displaystyle  %@@
Q}{\scriptstyle k}$}} S_i  %@@
 \equiv  \mbox{\raisebox{-1.1ex}{$\stackrel{\displaystyle  %@@
Q}{\scriptstyle k}$}}  %@@
\left(u_t^i-F^i \left(t,x, u, u_x, \ldots ,
u_{x}^{(k_i)}\right)\right)\Big\vert_{{\cal{M}}_p} =0,  \ i = 1,2,
\ldots, m, \ee
 where the manifold
${\cal{M}}_p$ is  one  $\{S_1=0,S_2=0, \dots, S_m=0, Q(u_{i_1})=0 ,
\dots, Q(u_{i_p})=0\}$ with  any given  numbers $\ i_1, \dots,i_p \,
( 1\leq p \leq  i_p \leq m)$.

\noindent {\textbf{ Definition 3.}} Operator (\ref{6}) is called the
$Q$-conditional symmetry (non-classical symmetry)  for  an evolution
system of the form (\ref{4}) if the following invariance conditions
are satisfied:
\be\label{10}  %@@
\mbox{\raisebox{-1.1ex}{$\stackrel{\displaystyle  %@@
Q}{\scriptstyle k}$}} S_i  %@@
 \equiv  \mbox{\raisebox{-1.1ex}{$\stackrel{\displaystyle  %@@
Q}{\scriptstyle k}$}}  %@@
\left(u_t^i-F^i \left(t,x, u, u_x, \ldots ,
u_{x}^{(k_i)}\right)\right)\Big\vert_{{\cal{M}}_m} =0,  \ i = 1,2,
\ldots, m, \ee
 where the manifold
${\cal{M}}_m$ is  one  $\{S_1=0,S_2=0, \dots, S_m=0, Q(u_{1})=0 ,
\dots, Q(u_{m})=0\}$.

Obviously, all three definitions coincide in the case of $m=1$, i.e.
a single evolution equation. If $m>1$ then one obtains a  hierarchy
of conditional symmetry operators. It is easily seen that
${\cal{M}}_m \subset {\cal{M}}_p \subset {\cal{M}}_1 \subset
{\cal{M}}$, hence, each Lie symmetry is automatically a
$Q$-conditional symmetry of the first and $p$-th types, while
$Q$-conditional symmetry of the first type is that of the $p$-th
type. From the formal point of view is enough to find all the
$Q$-conditional symmetry (non-classical symmetry) operators. Having
the full list of $Q$-conditional symmetries one may simply check
which of them is Lie symmetry or/and $Q$-conditional symmetry of the
$p$-th type.

%it is well-known that the main difficulty arising in search of
%$Q$-conditional symmetries is to solve so called system of
%DEs, which is nonlinear.

On the other hand,
 to construct any   $Q$-conditional symmetry  for  a
system of PDEs, one needs to solve another  nonlinear system, so
called system of DEs, which usually is much more complicated and
cumbersome. This problem arises even in the case of single linear
PDE and it was the reason why G.Bluman and J.Cole in their
well-known work \cite{bl-c} were unable to describe all
$Q$-conditional symmetries in explicit form even for the linear heat
equation. Thus, all three definitions are important from theoretical
and practical point of view. Notably, definition 1 should  be more
applicable  for solving systems of DEs when one examines a
multi-component systems containing three and more PDEs.

It should be stressed that definition 3 was only used in papers
\cite{lou-92}--\cite{ch-pli-08} devoted to the search
$Q$-conditional symmetries for the systems of PDEs. Moreover, to our
best knowledge, nobody has noted that several definitions producing
a  hierarchy  of conditional symmetry operators can be defined for
systems of  PDEs.

\begin{center}
{\textbf{3.  Conditional symmetries for the RD systems}}
\end{center}

Consider  the RD system  (\ref{1*}). According to the definitions
presented above, two types of conditional symmetry operators can be
derived: $Q$-conditional symmetry of the first type  and
$Q$-conditional symmetry of the second type. The second type
coincides with standard non-classical symmetry. It turns out that
systems of DEs corresponding to both types have essentially
different structure.

First of  all,  the RD system  (\ref{1*}) can be simplified if one
applies the Kirchhoff  substitution \be\label{3-1} u = \int
D^1(U)dU, \quad v = \int D^1(V)dV, \ee where $u(t,x)$ and $v(t,x)$
are new unknown functions. Hereafter we assume that there exist
inverse functions to those  arising in right-hand-sides of
(\ref{3-1}). Substituting (\ref{3-1}) into (\ref{1*}), one obtains
\be\label{1**}\ba{l} u_{xx}=d^1(u)u_t+C^1(u,v),
\\v_{xx}=d^2(v)v_t+C^2(u,v),\ea\ee
where the functions  $d^1,\ d^2$ and $C^1, C^2$ are uniquely defined
via $D^1,\ D^2$ and $F, G$, respectively.

Consider a conditional symmetry operator  of system (\ref{1*})
\begin{equation}
\label{3-2}
 {Q_*}= \xi_*^0(t,x,U,V)\partial_t+\xi_*^1(t,x,U,V)\partial_x+\eta_*^1(t,x,U,V)\partial_U+\eta_*^2(t,x,U,V)\partial_V,
\end{equation}
where $\xi_*^0, \xi_*^1,\eta_*^1$ and $\eta_*^2$ are to-be-found
functions.  The Kirchhoff  substitution transforms (\ref{3-2}) to
the same form
\begin{equation}
\label{3-3}
 {Q}= \xi^0(t,x,u,v)\partial_t+\xi^1(t,x,u,v)\partial_x+
 \eta^1(t,x,u,v)\partial_u+\eta^2(t,x,u,v)\partial_v.
\end{equation}
where  the operator coefficients are uniquely expressed via those
with stars using formulae (\ref{3-1}).

Let us use definition 1 to construct the  system of DEs for  finding
$Q$-conditional symmetry operators of the first type. According to
the definition the following invariance conditions must be
satisfied:
\be\label{3-4}\ba{l}  %@@
  \mbox{\raisebox{-1.1ex}{$\stackrel{\displaystyle  %@@
Q}{\scriptstyle 2}$}}  %@@
\left( d^1(u)u_t+C^1(u,v)- u_{xx} \right)\Big\vert_{{\cal{M}}_1}
=0,\\
\mbox{\raisebox{-1.1ex}{$\stackrel{\displaystyle  %@@
Q}{\scriptstyle 2}$}}  %@@
\left( d^2(v)v_t+C^2(u,v)- v_{xx} \right)\Big\vert_{{\cal{M}}_1} =0,
\ea \ee
 where the manifold
\be\label{3-4*}{\cal{M}}_1= \{S_1=0,S_2=0, Q(u)\equiv
\xi^0(t,x,u,v)u_t+\xi^1(t,x,u,v)u_x-
 \eta^1(t,x,u,v)=0\}. \ee Note that the
condition $Q(v)=0$  instead of $Q(u)=0$ can be also used, however,
it will lead to such conditional symmetry operators, which are
obtainable from those generated by the  invariance conditions
(\ref{3-4})--(\ref{3-4*}). In fact, the  simple renaming $u \to v, v
\to u$ (trivial discrete transformations) and the corresponding
coefficients' renaming preserve the form (\ref{1**}) and transform
the invariance criteria (\ref{3-4}) into that  with the condition
$Q(v)=0$ because of arbitrariness of the  functions $d^k$ and $C^k, \,
k=1,2 $.

{\textbf{ Remark 1.}} In the case of system (\ref{1**}) with the
%%precisely defined
fixed functions $d^k$ and $C^k, \, k=1,2 $ the discrete invariance
$u \to v, v \to u$ can be broken, so that definition 1 in two cases
should be examined.

Now we apply the rather standard procedure for obtaining system of
DEs, using the  invariance conditions (\ref{3-4}).
%% For the firsttime, to our best knowledge, its realization for the systems has
%%explicitly been presented  in \cite{lou-92}.
From the formal point of view, the procedure is the same as for Lie
symmetry search, however, three (not two !) different derivatives,
say $u_{xx}, v_{xx}, u_t $, can be excluded using the manifold
${\cal{M}}_1$.
%% \cite{fss, bl-c, ch-se-03},
After rather cumbersome calculations, one arrives at the nonlinear
system of DEs
\begin{equation}\label{3-5}\ba{l}
1)\ \medskip \xi^0_{x}=\xi^0_{u}=\xi^0_{v} =\xi^1_{u}=\xi^1_{v}=0,
\\ 2)\ \medskip \eta^1_{v}=\eta^1_{uu}=\eta^2_{uu}=
\eta^2_{vv}=\eta^2_{uv}=0,
\\ 3)\ \medskip \xi^1\eta^2_u(d^2-d^1)+2\xi^0\eta^2_{xu}=0,
\\ 4)\ \medskip  (\xi_t^0\xi^1-\xi^0\xi^1_t-2\xi^1\xi^1_x)d^1-\xi^1\eta^1d^1_u
-2\xi^0\eta^1_{xu}+\xi^0\xi^1_{xx}=0,
\\ 5)\ \medskip  (2\xi^1_x-\xi^0_t)d^2 +\eta^2d^2_v =0, \\
6)\ \medskip \xi^1_td^2+2\eta^2_{xv}-\xi^1_{xx}=0,
\\ 7)\ \frac{\eta^1}{\xi^0} \eta^1d^1_u+(\eta^1_t+2\xi^1_x\frac{\eta^1}{\xi^0}
- \xi^0_t\frac{\eta^1}{\xi^0})d^1 -\eta^1_{xx}+
%%\\ \medskip
 \eta^1C^1_u+\eta^2C^1_v+(2\xi^1_x-\eta^1_u)C^1=0,
\\ 8)\  \medskip (\eta^2_t+\frac{\eta^1}{\xi^0}\eta^2_u)d^2-\frac{\eta^1}{\xi^0}\eta^2_u d^1-\eta^2_{xx}
+\eta^1C^2_u+\eta^2C^2_v-\eta^2_uC^1+ (2\xi^1_x-\eta^2_v)C^2=0,
\ea\end{equation} if $\xi^0\not=0$ and $d^1(u)\not=d^2(v)$ (the
special cases $\xi^0=0$ and  $d^1(u)=d^2(v) =const$ can be
treated in the similar way, however,  the results omitted here  to
avoid cumbersome formulae).

{\it An  analysis of system (\ref{3-5}) shows that finding of
$Q$-conditional symmetry of the first type  cannot be reduced to the
operators (\ref{3-3}) with $\xi^0=1$.  This is in contradiction to
the well-known fact occurring  in the case of  single evolution
equations.} Indeed, if one assumes that system (\ref{3-5}) is
locally equivalent to that  with $\xi^0=1$ then equation 5) takes
the form
 \[ 2\xi^1_xd^2 +\eta^2d^2_v =0, \]
 therefore the restriction $\xi^1_x=0$ is obtained provided
 $d^2=const \not=0$.
   On the other hand, there are several RD
 systems, which are invariant under the Lie symmetry operators of
 the form (see \cite{ch-king}, table 3)
 \[2t\partial_t +x\partial_x +\alpha_1 u\partial_u +
 \alpha_2 v\partial_v, \]
 (here  $\alpha_1$ and $\alpha_2 $ are
 constants), i.e., those with $\xi^1_x\not=0$. Obviously, such operators
 %of the form \[\partial_t +\frac{x}{2t}\partial_x +\alpha_1 \frac{u}{2t}\partial_u +
 %\alpha_2 \frac{v}{2t}\partial_v, \]
 will be lost if one assumes $\xi^0=1$ in the very beginning.

Now we apply definition 3 to construct the  system of DEs for
finding $Q$-conditional symmetry (non-classical symmetry) operators.
Thus, the  invariance conditions
\be\label{3-6}\ba{l}  %@@
  \mbox{\raisebox{-1.1ex}{$\stackrel{\displaystyle  %@@
Q}{\scriptstyle 2}$}}  %@@
\left( d^1(u)u_t+C^1(u,v)- u_{xx} \right)\Big\vert_{{\cal{M}}_2}
=0,\\
\mbox{\raisebox{-1.1ex}{$\stackrel{\displaystyle  %@@
Q}{\scriptstyle 2}$}}  %@@
\left( d^2(v)v_t+C^2(u,v)- v_{xx} \right)\Big\vert_{{\cal{M}}_2} =0
\ea \ee must be satisfied. Here  the manifold ${\cal{M}}_2=
\{S_1=0,S_2=0, Q(u)=0, Q(v)=0  \}$, so that   four different
derivatives, say $u_{xx}, v_{xx}, u_t, v_t $, can be excluded in
this case. Finally, the system of DEs
\begin{equation}\label{3-7}\ba{l}
1)\ \medskip \xi^0_{x}=\xi^0_{u}=\xi^0_{v}
=\xi^1_{uu}=\xi^1_{vu}=\xi^1_{vv}=0,
\\ 2)\ \medskip \eta^1_{vv}=\eta^2_{uu}=0,
\\ 3)\ \medskip 2\frac{\xi^1}{\xi^0}\xi^1_u d^1+\eta^1_{uu}-2\xi^1_{xu}=0,
\\ 4)\ \medskip 2\frac{\xi^1}{\xi^0}\xi^1_v d^2+\eta^2_{vv}-2\xi^1_{xv}=0,
\\ 5)\ \medskip \frac{\xi^1}{\xi^0}\xi^1_v(d^1+d^2)+2\eta^1_{uv}-2\xi^1_{xv}=0,
\\ 6)\ \medskip \frac{\xi^1}{\xi^0}\xi^1_u(d^1+d^2)+2\eta^2_{uv}-2\xi^1_{xu}=0,
\\ 7)\ \medskip \frac{\xi^1}{\xi^0}\eta^1_v(d^1-d^2)-2\frac{\eta^1}{\xi^0}\xi^1_vd^1+2\eta^1_{xv}-2\xi^1_v C^1=0,
\\ 8)\ \medskip \frac{\xi^1}{\xi^0}\eta^2_u(d^2-d^1)-2\frac{\eta^2}{\xi^0}\xi^1_ud^2+2\eta^2_{xu}-2\xi^1_u C^2=0,
\\ 9)\ \medskip (\frac{\xi^1}{\xi^0}\xi^0_t+2\frac{\eta^1}{\xi^0}\xi^1_u-\xi^1_t-\frac{\eta^2}{\xi^0}\xi^1_v-2\frac{\xi^1}{\xi^0}\xi^1_x)d^1-\frac{\xi^1}{\xi^0}\eta^1d^1_u+
\\ \medskip \qquad +\frac{\eta^2}{\xi^0}\xi^1_vd^2-2\eta^1_{xu}+\xi^1_{xx}+3\xi^1_uC^1+\xi^1_vC^2=0,
\\ 10)\ \medskip (\frac{\xi^1}{\xi^0}\xi^0_t+2\frac{\eta^2}{\xi^0}\xi^1_v-\xi^1_t-\frac{\eta^1}{\xi^0}\xi^1_u-2\frac{\xi^1}{\xi^0}\xi^1_x)d^2-\frac{\xi^1}{\xi^0}\eta^2d^2_v+
\\ \medskip \qquad +\frac{\eta^1}{\xi^0}\xi^1_u d^1-2\eta^2_{xv}+\xi^1_{xx}+3\xi^1_vC^2+\xi^1_uC^1=0,
\\ 11)\ (\eta^1_t-\frac{\eta^1}{\xi^0}\xi^0_t+\frac{\eta^2}{\xi^0}\eta^1_v+2\frac{\eta^1}{\xi^0}\xi^1_x)d^1+\frac{\eta^1}{\xi^0}\eta^1d^1_u-\frac{\eta^2}{\xi^0}\eta^1_vd^2-\eta^1_{xx}
\\ \medskip \qquad + \eta^1C^1_u+\eta^2C^1_v+(2\xi^1_x-\eta^1_u)C^1-\eta^1_vC^2=0,
\\ 12)\  (\eta^2_t-\frac{\eta^2}{\xi^0}\xi^0_t+\frac{\eta^1}{\xi^0}\eta^2_u+2\frac{\eta^2}{\xi^0}\xi^1_x)d^2+\frac{\eta^2}{\xi^0}\eta^2d^2_v-\frac{\eta^1}{\xi^0}\eta^2_ud^1-\eta^2_{xx}
\\ \medskip \qquad +\eta^1C^2_u+\eta^2C^2_v-\eta^2_uC^1+(2\xi^1_x-\eta^2_v)C^2=0
\ea\end{equation} is obtained if $\xi^0\not=0$ (the special case
$\xi^0=0$ can be treated in the quite  similar way).

{\it An  analysis of system (\ref{3-7}) shows that finding
$Q$-conditional symmetry operators for  the RD system  (\ref{1*})
can be reduced to the operators (\ref{3-3}) with $\xi^0=1$ provided
$\xi^0\not=0$.} In fact, the substitution \be\label{3-8}
\xi_*=\frac{\xi^1}{\xi^0}, \, \eta^k_*=\frac{\eta^k}{\xi^0}, k=1,2
\ee reduces  system (\ref{3-7}) to the form (the stars next to the
functions $ \xi$ and $\eta^k$ are skipped):

\begin{equation}\label{3-9}\ba{l}
1)\ \medskip \xi_{uu}=\xi_{vv}=\xi_{uv}=0,
\\ 2)\ \medskip \eta^1_{vv}=\eta^2_{uu}=0,
\\ 3)\ \medskip 2\xi\xi_u d^1+\eta^1_{uu}-2\xi_{xu}=0,
\\ 4)\ \medskip 2\xi\xi_v d^2+\eta^2_{vv}-2\xi_{xv}=0,
\\ 5)\ \medskip \xi\xi_v(d^1+d^2)+2\eta^1_{uv}-2\xi_{xv}=0,
\\ 6)\ \medskip \xi\xi_u(d^1+d^2)+2\eta^2_{uv}-2\xi_{xu}=0,
\\ 7)\ \medskip \xi\eta^1_v(d^1-d^2)-2\xi_v\eta^1d^1+2\eta^1_{xv}-2\xi_v C^1=0,
\\ 8)\ \medskip \xi\eta^2_u(d^2-d^1)-2\xi_u\eta^2d^2+2\eta^2_{xu}-2\xi_u C^2=0,
\\ 9)\  (2\xi_u\eta^1-\xi_t-\xi_v\eta^2-2\xi\xi_x)d^1-\xi\eta^1d^1_u+
\\ \medskip \qquad+\xi_v\eta^2d^2-2\eta^1_{xu}+\xi_{xx}+3\xi_uC^1+\xi_vC^2=0,
\\ 10)\  (2\xi_v\eta^2-\xi_t-\xi_u\eta^1-2\xi\xi_x)d^2-\xi\eta^2d^2_v+
\\ \medskip \qquad+\xi_u\eta^1d^1-2\eta^2_{xv}+\xi_{xx}+3\xi_vC^2+\xi_uC^1=0,
\\ 11)\ (\eta^1_t+\eta^2\eta^1_v+2\xi_x\eta^1)d^1+(\eta^1)^2d^1_u-\eta^2\eta^1_vd^2-\eta^1_{xx}
\\ \medskip \qquad+ \eta^1C^1_u+\eta^2C^1_v+(2\xi_x-\eta^1_u)C^1-\eta^1_vC^2=0,
\\ 12)\  (\eta^2_t+\eta^1\eta^2_u+2\xi_x\eta^2)d^2+(\eta^2)^2d^2_v-\eta^1\eta^2_ud^1-\eta^2_{xx}
\\ \medskip \qquad+\eta^1C^2_u+\eta^2C^2_v-\eta^2_uC^1+(2\xi_x-\eta^2_v)C^2=0
\ea\end{equation} and $\xi^0$ is an arbitrary smooth function of the
time variable. Now we realize that system (\ref{3-9}) is nothing
else but the system of DEs to find the $Q$-conditional symmetry
operators (\ref{3-3}) with $\xi^0=1$. Thus,
  to find all   $Q$-conditional symmetries of the RD system
(\ref{1*}) one needs to solve only the particular case of system
(\ref{3-7}), i.e., system (\ref{3-9}).

\medskip

{\textbf{ Remark 2.}} System (\ref{3-9}) with
$d^1(u)=u^m,d^2(v)=v^n$ coincides with the system of DEs obtained
and analyzed in the recent paper \cite{ch-pli-08}.

\medskip

{\textbf{ Remark 3.}}  System of DEs to search Lie symmetries of RD
systems with non-constant diffusivities (see equations (9)--(13) in
\cite{ch-king4}) differs essentially from   systems  (\ref{3-5}) and
(\ref{3-7}). In contrary to  systems  of DEs for $Q$-conditional
symmetries,  one for Lie symmetries contains the subsystem of
equations
\[\xi^0_{u}=\xi^0_{v} =\xi^1_{u}=\xi^1_{v}=\eta^1_{v}=\eta^2_{u}=0 \]
what essentially simplifies its solving.

\medskip

Of course, system of DEs (\ref{3-5}) can be derived from system
(\ref{3-7}) because each $Q$-conditional symmetry of the first type
is automatically a $Q$-conditional symmetry (but not vice  versa !).
However,  system  (\ref{3-7}) is much complicated than (\ref{3-5})
and its solving is a difficult problem even in particular cases,
e.g., $d^1(u)=u^m,d^2(v)=v^n$ \cite{ch-pli-08}. Thus, we believe
that $Q$-conditional symmetries of the first type can  be found much
easier for many  nonlinear RD systems   (\ref{1*}) with
correctly-specified coefficients, arising in applications.

Now  we present an example, which highlights when $Q$-conditional
symmetry (non-classical symmetry) is or is  not  of the first type.

{\textbf{ Example 1.}} Consider system (\ref{3-7}) assuming
\begin{equation}\label{3-10}
\xi^1= \eta^1_{v}=\eta^2_{u}=0.\end{equation}
 In this case, the system can be immediately  simplified
 and one obtains only two equations
 \begin{equation}\label{3-11}\ba{l}
 (\eta^1_t-\frac{\eta^1}{\xi^0}\xi^0_t)d^1+\frac{\eta^1}{\xi^0}\eta^1d^1_u-\eta^1_{xx}
+ \eta^1C^1_u+\eta^2C^1_v-\eta^1_uC^1=0,
\\   (\eta^2_t-\frac{\eta^2}{\xi^0}\xi^0_t)d^2+\frac{\eta^2}{\xi^0}\eta^2d^2_v-\eta^2_{xx}
 +\eta^1C^2_u+\eta^2C^2_v-\eta^2_vC^2=0
\ea\end{equation} to find the functions
 \be\label{3-12}\ba{l}
\xi^0=c(t),\\
\eta^1=r^1(t)u+p^1(t,x),\\
\eta^2=r^2(t)v+p^2(t,x),\ea\ee where the functions in
right-hand-sides should   be determined.

System (\ref{3-5}) with restrictions  (\ref{3-10}) takes the form
\begin{equation}\label{3-13}\ba{l}
 (\eta^1_t-\frac{\eta^1}{\xi^0}\xi^0_t)d^1+\frac{\eta^1}{\xi^0}\eta^1d^1_u-\eta^1_{xx}
+ \eta^1C^1_u+\eta^2C^1_v-\eta^1_uC^1=0,
\\   \eta^2_td^2-\eta^2_{xx}
 +\eta^1C^2_u+\eta^2C^2_v-\eta^2_vC^2=0,
 \\ \xi^0_td^2 = \eta^2d^2_v,
\ea\end{equation} where the unknown functions $\xi^0 $ and $\eta^k,
k=1,2$ must be of the form (\ref{3-12}).

Now one notes that systems (\ref{3-11}) and (\ref{3-13}) are
equivalent if $\eta^2d^2_v =0$ (we remind the reader that
substitution (\ref{3-8}) reduces system (\ref{3-7}) to that  with
$\xi^0=1$). Thus, we arrive at the
%%conclusion:
statement: {\it each  $Q$-conditional symmetry operator (\ref{3-3})
of the RD system (\ref{1**}) with $d^2(v)= const $ (or $d^1(u)=
const $)
 is equivalent (up to multiplier $c(t)\not=0$) to  a
$Q$-conditional symmetry operator of the first type if restrictions
(\ref{3-10}) take place. Moreover, the statement  is valid for all
systems of the form (\ref{1**}) under the additional restriction
$\eta^2=0$ (or $\eta^1=0$).}

Taking into account the Kirchhoff  substitution,  the same statement
can be formulated for RD system (\ref{1*}) because substitution
(\ref{3-1}) preserves restrictions (\ref{3-10}).

Finally, we analyze briefly the RD system  (\ref{3}) (the cases
$l=k=0$ and $l=k=-1$ are excluded) using paper  table 1
\cite{ch-pli-08}, which presents all possible $Q$-conditional
symmetry (non-classical symmetry) operators of (\ref{3}) under
restrictions (\ref{3-10}). There are five different cases according
to theorem 1 \cite{ch-pli-08}. Using the statement obtained, we
conclude that the $Q$-conditional symmetry operator arising in the
first case of table 1  is simultaneously a
$Q$-conditional symmetry operator of the first type while the
operator arising in the third case is not because $l=k=-\frac{1}{2}
\not=0$ and $\eta^1\eta^2\not=0$. Other three operators of
$Q$-conditional symmetry from table 1  are those of
$Q$-conditional symmetry of the first type only under additional
restrictions on coefficients of the corresponding RD systems. These
restrictions can be easily derived using the statement formulated
above. For example, the $Q$-conditional symmetry operator listed the
second  case of table 1  is that of the first type
under the  restrictions either $\lambda_1=0$ or $l=0$ (if
$\lambda_1=l=0$ then it  is the Lie symmetry operator of the
corresponding RD system).

\vspace{5mm}

{\bf Table 1.   $Q$-conditional symmetries of the RD system
(\ref{3})} \footnotesize
\begin{center}
\begin{tabular}{|c|c|c|c|c|}
\hline
no & RD systems of the form (\ref{1*}) &Q-conditional operators& Restrictions    \\
%% &   F and G & operators &conditions   \\
\hline
 &&& \\
1. & $U_t=(U^kU_x)_x+f(U^{k+1}),$ &  $\partial_t+2p(x)V^{1\over2}\partial_V$ & $p_{xx}=(p)^2+\lambda p,$ \\
  &$   V_t=(V^{-\frac{1}{2}}V_x)_x -2 \lambda V^{1\over2}+g(U^{k+1})$& &$  p\ne0$ \\

\hline
 &&& \\

2. & $U_t=(U^kU_x)_x+\lambda_1 U^{-k}+ f(U^{k+1}-\alpha V^{l+1}), $ & $ \partial_t+\lambda_1U^{-k}\partial_U+$ &$\alpha=\frac{\lambda_1(k+1)}{\lambda_2(l+1)},\ \lambda_2\ne0$  \\
  &$  V_t=(V^{l}V_x)_x + \lambda_2 V^{-l}+g(U^{k+1}-\alpha V^{l+1}) $   & $+\lambda_2 V^{-l}\partial_V$ &$
\lbd_1^2+l^2\ne0.$\\
\hline
 &&& \\

3. & $U_t=(U^{-{1\over2}}U_x)_x-2\lambda U^{1\over2}+ f(U^{1\over2}-
V^{1\over2}), $
& $\partial_t+$ & $p_{xx}=(p)^2+\lambda p, $ \\
  &$ V_t=(V^{-\frac{1}{2}}V_x)_x-2 \lambda V^{1\over2}+g(U^{1\over2}- V^{1\over2}) $
   & $+2p(x)(U^{1\over2}\partial_U+V^{1\over2}\partial_V)$ & $ p\ne0 $ \\
\hline
 &&& \\

4.&  $ U_t=(U^kU_x)_x+\lambda_1 U^{-k}+ f(\omega), $ &$\partial_t+\lambda_1U^{-k}\partial_U+$ & $\omega=\frac{\exp U^{k+1}}{(V^{l+1}-\lambda_3)^{\frac{\lambda_1(k+1)}{\lbd_2(l+1)}}},\ \lbd_2\ne0,$ \\
 & $V_t=(V^lV_x)_x+(V^{l+1}-\lambda_3)(g(\omega)+\lambda_2 V^{-l}) $
 &$+\lambda_2(V-\lambda_3V^{-l})\partial_V$&$either\ \lbd_1^2+\lbd_3^2\ne0\ or $\\
&$  $& &$  \lbd_3^2+k^2\ne0\ or \lbd_1^2+l^2\ne0$\\ \hline
 &&& \\

5.&  $ U_t=(U^kU_x)_x+( U^{k+1}-\lambda_1)(f(\omega)+\lambda_2
U^{-k})  $ &
$\partial_t+\lambda_2(U-\lambda_1U^{-k})\partial_U+$  & $\omega=\frac{U^{k+1}-\lambda_1}{(V^{l+1}-\lambda_3)^{{\lambda_2(k+1)}\over{\lambda_4(l+1)}}},\ \lambda_2 \lambda_4\ne0,$ \\
 & $ V_t=(V^lV_x)_x+(V^{l+1}-\lambda_3)(g(\omega)+\lambda_4 V^{-l})$ &$+\lambda_4(V-\lambda_3
V^{-l})\partial_V$ & $either\ \lbd_1^2+\lbd_3^2\ne0,\ $  \\
&$  $& &$ or \ \lbd_3^2+k^2\ne0\ or \ \lbd_1^2+l^2\ne0$\\

\hline
\end{tabular}
\end{center}
\normalsize

\begin{center}
{\textbf{4.  Conditional symmetries of the first type for the
diffusive
 Lotka-Volterra system }}
\end{center}

Here we  consider the diffusive Lotka-Volterra
 (DLV) system
\be\label{2-rev}\ba{l}
\lbd_1 u_t =  u_{xx}+u(a_1+b_1u+c_1v),\\ %@@
\lbd_2 v_t = v_{xx}+ v(a_2+b_2u+c_2v),  %@@
%% \ea  %@@
\ea\ee  %@@
which is  the most common particular case of reaction-diffusion (RD)
system (\ref{1**}). The parameters $\lbd_k, a_k, b_k,$ and $ c_k$
($k=1,2$)  are assumed to be arbitrary non-zero constants.

 System (\ref{1*}) is the simplest generalization
of the classical Lotka-Volterra  system that takes into account the
diffusion process for interacting species (see terms  $u_{xx}$ and
$v_{xx}$). Nevertheless the classical Lotka-Volterra  system  was
introduced by A.J.Lotka and V.Volterra more than 80 years ago, its
natural generalization (\ref{2-rev}) is still  studied because this
is one of the most important mathematical models.   Lie symmetries
of (\ref{2-rev}) have been completely described in \cite{ch-du-04}
(note those can be extracted from more general results presented in
\cite{ch-king,ch-king2}).

\bt In the case  $\lambda_1\neq\lambda_2$,    DLV system
(\ref{2-rev}) is invariant under $Q$-conditional operators of the
first type if and only if the corresponding system   and
$Q$-conditional symmetries (up to local transformations
$u\rightarrow bu, \ v\rightarrow cv, \ bc \not=0 $ ) have the forms \be\label{4*}
\ba{l}
 \quad \ \lbd_1 u_t = u_{xx}+u(a_1+u+v),\\
 \quad \ \lbd_2 v_t = v_{xx}+ v(a_2+u+v),\quad a_1\neq a_2, \\
\ea \ee
\be\label{4-0}\quad{Q_1}=(\lbd_1-\lbd_2)\p_t+(a_1-a_2)u(\p_u-\p_v),
 \ee \be\label{4-1}\quad
{Q}_2=(\lbd_1-\lbd_2)\p_t+(a_1-a_2)v(\p_v - \p_u).\ee Each other
$Q$-conditional operator of the first type coincides with the Lie
symmetry operator.

In the case  $\lambda_1=\lambda_2$,    DLV system (\ref{2-rev}) is
invariant  only under such $Q$-conditional operators of the first
type, which coincide with the Lie symmetry operators.
  \et

\textbf{Proof.} Proof  is based on solving the system of DEs (\ref{3-5})
with $d^k= \lbd_k,$ $C^1=u(a_1+b_1u+c_1v)$ and
$C^2=v(a_2+b_2u+c_2v)$. After rather simple calculations, one
obtains that  DLV system (\ref{2-rev}) is invariant  only under a
$Q$-conditional operator of the first type only under the
coefficient restrictions
\[ \lambda_1\neq\lambda_2, \quad b_1=b_2=b, \quad c_1=c_2=c. \]
In this case  the  transformations $u\rightarrow bu, \ v\rightarrow
cv $ reduce the DLV system  to the form (\ref{4*}). The
corresponding operator takes the form (\ref{4-0}).

To find operator (\ref{4-1}) one needs to proceed the same algorithm
but with the invariance criteria (\ref{3-4}) for the manifold
\be\label{3-4**}{\cal{M}}_1= \{S_1=0,S_2=0, Q(v)\equiv
\xi^0(t,x,u,v)v_t+\xi^1(t,x,u,v)v_x-
 \eta^2(t,x,u,v)=0\}, \ee
 i.e., to use definition 1 with $i_1=2$.

\medskip

\textbf{Remark 4.} Theorem 1  gives {\it a complete description of
$Q$-conditional symmetries of the first type in explicit form} for
 DLV system (\ref{2-rev}). It turns out that all possible    $Q$-conditional symmetries of the
second type    cannot
 be derived in a similar way. In fact, the system of DEs (\ref{3-9}) with $d^k=
\lbd_k,$ $C^1=u(a_1+b_1u+c_1v)$ and $C^2=v(a_2+b_2u+c_2v)$ has much
more complicated structure and can be solved only under some
additional assumptions.

\medskip

The $Q$-conditional symmetry operators  (\ref{4-0}) and (\ref{4-1})
can be applied for finding exact solutions of DLV system (\ref{4*})
using the same algorithm  as for classical Lie symmetries. Since
both operators have the same structure we examine only the second
one.
 Thus, the corresponding ansatz reads 
 \be\label{4-3}
\ba{l}
 u(t,x)=\varphi_1(x)-\varphi_2(x)\exp(\frac{a_1-a_2}{\lambda_1-\lambda_2}t),\\
 v(t,x)=\varphi_2(x)\exp(\frac{a_1-a_2}{\lambda_1-\lambda_2}t)
\ea \ee
and  reduces system (\ref{4*}) to the ODE system
 \be\label{4-4}
\ba{l}\varphi''_1+\varphi^2_1+a_1\varphi_1=0, \\
\varphi''_2+\frac{a_2\lambda_1-a_1\lambda_2}{\lambda_1-\lambda_2}\varphi_2+
\varphi_1\varphi_2=0, \ea\ee  where $\varphi_1(x)$ and
$\varphi_2(x)$ are to be determined functions.

The general solution of this nonlinear ODE system cannot  be found
in an explicit form. However,  we constructed its particular
solutions, which lead to interesting solutions of the corresponding
DLV system. In fact, the first ODE in (\ref{4-1}) possesses the
non-zero steady state solution $\varphi_1=-a_1$. Substituting this
solution into the second ODE in (\ref{4-1}), one arrives at the
linear ODE \be\label{105}
 \varphi''_2-\beta \lambda_1\varphi_2=0,
 \ee
where $\beta=\frac{a_1-a_2}{\lambda_1-\lambda_2}\not=0.$ Thus, using
the general solution of  (\ref{105}) and ansatz (\ref{4-3}),  we
obtain   two families of exact solutions of the DLV system
(\ref{4*}): \be\label{106} \ba{l}
\medskip
u(t,x)=-a_1+\frac{1}{a_2-a_1}\big(C_1\exp(\sqrt{\beta\lambda_1}x)+C_2\exp(-\sqrt{\beta\lambda_1}x)\big)e^{\beta t},\\
v(t,x)=\frac{1}{a_1-a_2}\big(C_1\exp(\sqrt{\beta\lambda_1}x)+C_2\exp(-\sqrt{\beta\lambda_1}x)\big)e^{\beta
t}, \ea \ee if $\beta>0,$  and   \be\label{107} \ba{l}
\medskip
u(t,x)=-a_1+\frac{1}{a_2-a_1}\big(C_1\cos(\sqrt{-\beta\lambda_1}x)+C_2\sin(\sqrt{-\beta\lambda_1}x)\big)e^{\beta t},\\
v(t,x)=\frac{1}{a_1-a_2}\big(C_1\cos(\sqrt{-\beta\lambda_1}x)+C_2\sin(\sqrt{-\beta\lambda_1}x)\big)e^{\beta
t}, \ea \ee if  $\beta<0$ (hereafter  $C_1,  C_2$ are arbitrary
constants).

It should be noted that  solutions (\ref{106}) and (\ref{107})
cannot be constructed using Lie symmetries. In fact, DLV system
(\ref{4*})   admits only the trivial Lie symmetry \cite{ch-du-04},
hence, the  plane wave solutions only  can be found by Lie symmetry
reductions (the reader may find examples of such solutions in
\cite{rod-mimura-2000} and  \cite{ch-du-04}). Obviously, solutions
(\ref{106}) and (\ref{107}) possess  more complicated structures.

\vspace{0.5cm}

\textbf{Example 2.} Consider solution  (\ref{107}) with  $C_1=0$.
 Using the substitution  $u\rightarrow -bu, \ v\rightarrow -cv  \ (b>0,\ c>0),$
 one transforms DLV system  (\ref{4*}) to the standard system describing the competition of
two species (see, e.g., \cite {mur2}):
 \be\label{136} \ba{l}
 \lbd_1 u_t = u_{xx}+u(a_1-bu-cv),\\
 \lbd_2 v_t = v_{xx}+ v(a_2-bu-cv)\ea \ee
and   solution  (\ref{107}) to the form \be\label{134} \ba{l}
\medskip
u(t,x)=\frac{a_1}{b}+\frac{1}{(a_1-a_2)b} \ C_2\sin(\sqrt{-\beta\lambda_1}x)e^{\beta t},\\
v(t,x)=\frac{1}{(a_2-a_1)c} \
C_2\sin(\sqrt{-\beta\lambda_1}x)e^{\beta t}, \ea \ee where the
coefficient restrictions $\beta \equiv
\frac{a_1-a_2}{\lambda_1-\lambda_2}<0, \ a_1>0,\ a_2>0$ are assumed.

One notes that  this solution  satisfies the constant Dirichlet
conditions
\be\label{134b} \ba{l} x=0:\  u= \frac{a_1}{b}, \ v=0,  \\
x=\frac{\pi}{\sqrt{-\beta\lambda_1}}:\  u= \frac{a_1}{b}, \ v=0, \ea
\ee in the domain  $\Omega=\{ (t,x) \in (0,+ \infty )\times
\Bigl(0,\frac{\pi}{\sqrt{-\beta\lambda_1}}\Bigr)\}. $ Moreover,
solution (\ref {134})
 has the time asymptotic
 \be\label{135} (u,\ v)\rightarrow (\frac{a_1}{b}, \ 0), \quad
t\rightarrow +\infty.\ee

 Thus, this solution  describes  the
competition between  two species  when   the species $u$ eventually
%%wipes out
dominate  while  the species $v$ die.

\begin{center}
{\textbf{5. Conclusions}}
\end{center}

In this paper, new  definitions of $Q$-conditional
  symmetry for systems of PDEs  are presented,
  which generalize the standard notation of
  non-classical (conditional) symmetry.
  It is shown that different types of $Q$-conditional
  symmetry  generate  a  hierarchy  of conditional symmetry
  operators. Since  conditional symmetries can be applied for
finding exact solutions of the relevant equations, which are not
obtainable by the classical Lie method,  we  demonstrated this for
constructing new exact solutions of the diffusive
 Lotka-Volterra system.

Systems of DEs to find
 $Q$-conditional symmetries of two types for
 the nonlinear RD system (\ref{1*}) are constructed.
 The case of conditional invariance under   operator (\ref{3-2}) with
the coefficients $\xi^1=0$, $\eta^1_*(t,x,U)$ and $\eta^2_*(t,x,V)$
is analyzed in details. Using the recent paper \cite{ch-pli-08},
  we established that there are
 the nonlinear RD systems of five types, which
possess $Q$-conditional symmetry (non-classical symmetry) operators
satisfying definition 3. However, only one of them  is
simultaneously a $Q$-conditional symmetry operator of the first
type.
%% according to definition 1.

Definitions 1--3 can be straightforwardly extended on an arbitrary
$m$-component system of PDEs  presented in a 'canonical' form (the
system has a simplest form and there are no any non-trivial
differential consequences). However, new difficulties may arise
because differential consequences of additional conditions
 \be\label{5-1} Q(u_{1})=0 , \dots, Q(u_{m})=0,
\ee generated by the conditional symmetry operator (\ref{6}). In
fact, there are examples of single  hyperbolic PDEs when the set of
$Q$-conditional symmetry operators can be extended if one takes into
account  such differential consequences. For example, this occurs in
the case of nonlinear hyperbolic equation $u_{tt}=u u_{xx}$. The
reader may easily check that there is a much wider set of
$Q$-conditional symmetry operators in \cite{fss} (Chapter 5) than it
was found in \cite{pol-za-04} (Supplement 7) neglecting differential
consequences. It is the reason why there is the definition of
$Q$-conditional invariance (non-classical invariance) for a $k$-th
order single PDE, which requires to take into account all
differential consequences of the additional condition $Q(u)=0$ up to
the order $k$ (see, e.g.,\cite{zh-tsy-pop}). From this point of
view, definitions 1--3 can be formally extended by using
differential consequences of (\ref{5-1}) up to the order $k = \max
\{k_i, \ i = 1, \ldots, m \}$. However, we believe that such
extensions produce much more cumbersome formulae but don't lead to
any new operators for evolution system (\ref{4}) because all the
above mentioned differential consequences contain the mixed and/or
high-order time derivatives, $u_{tx}, u_{tt}, u_{ttx},\ldots$, which
are absent in system (\ref{4}). For example, if one applies the
`amended' definition 1 with ${\cal{M}}^*_1=\{S_1=0,S_2=0, Q(u)=0,
\partial_t Q(u)=0, \partial_x Q(u)=0\}$ (instead of (\ref{3-4*}))
to find $Q$-conditional symmetry operators of the first type for the
RD system (\ref{1**}) then the system of DEs (\ref{3-5}) is exactly
obtained but nothing more.

\medskip
\begin{center}
\textbf{ Acknowledgment} \end{center}

  The author is grateful
to the unknown   referees for the useful comments.

\end{document}